\documentclass{optica-article}

\journal{opticajournal} 

\articletype{Research Article}

\usepackage{lineno}
\usepackage{subfigure}
\usepackage{epstopdf}
\usepackage{bm}

\begin{document}

\title{Continuous-variable quantum key distribution network based on entangled states of optical frequency combs}

\author{Hai Zhong,\authormark{1} Qianqian Hu,\authormark{1} Zhiyue Zuo,\authormark{2,*} Zhipeng Wang,\authormark{2} Duan Huang,\authormark{3} and Ying Guo\authormark{2,4}}

\address{\authormark{1}School of Computer Science \& Technology, Changsha University of Science and Technology, Changsha 410114, China \\
\authormark{2}School of Automation, Central South University, Changsha 410083, China\\
\authormark{3}School of Electronic Information, Central South University, Changsha 410083, China\\
\authormark{4}School of Computer Science, Beijing University of Posts and Telecommunications, Beijing 100876, China}

\email{\authormark{*}zuozuo@csu.edu.cn} 


\begin{abstract*}
Continuous-variable quantum key distribution (CVQKD) features a high key rate and compatibility with classical optical communication. Developing expandable and efficient CVQKD networks will promote the deployment of large-scale quantum communication networks in the future. This paper proposes a CVQKD network based on the entangled states of an optical frequency comb. This scheme generates Einstein-Podolsky-Rosen entangled states with a frequency comb structure through the process of a type-II optical parametric oscillator. By combining with the scheme of entanglement in the middle, a fully connected CVQKD network capable of distributing secret keys simultaneously can be formed. We analyze the security of the system in the asymptotic case. Simulation results show that under commendable controlling of system loss and noise, the proposed scheme is feasible for deploying a short-distance fully connected CVQKD network. Loss will be the main factor limiting the system's performance. The proposed scheme provides new ideas for a multi-user fully connected CVQKD network.
\end{abstract*}

\section{Introduction}
Quantum key distribution (QKD)~\cite{Gisin02RMP, Pira20AOP} enables two legitimate parties, Alice and Bob, to establish a shared secure key through an insecure quantum channel. It can be divided into two major branches: discrete variable (DV) QKD and continuous variable (CV) QKD. CVQKD encodes information onto the two quadratures of the optical field and extracts information using coherent detection, making it feature a high secret key rate and compatibility with classical optical communication. Thus, it has advantages in short-distance high-speed QKD applications. After more than two decades of development, CVQKD has made remarkable progress, especially for the protocol of Gaussian-modulated coherent states \cite{GG02}. Theoretically, the unconditional security of CVQKD has been fully proven~\cite{Leve15, Leve17, Denys21}, and many researchers have also proposed multiple performance improvement schemes~\cite{Guo17PRA, Guo19PRA, Liao20NJP}. Experimentally, high-speed and long-distance CVQKD systems have been implemented~\cite{Huang15OE, Huang16SR, Zhang20PRL}, and multiple field test experiments have also demonstrated the feasibility of deploying CVQKD in existing optical communication networks~\cite{Foss09NJP, Huang16OL}. Currently, point-to-point one-way CVQKD systems are gradually moving towards engineering, practical application, and networking.

However, a point-to-point CVQKD system cannot meet the future large-scale key distribution requirements. In the face of the scenario of simultaneous communication in multi-user networking, the development of a multi-user fully connected network technology for CVQKD is imperative. Based on entanglement distribution, the fully connected and simultaneously communicating DVQKD network has been experimentally verified~\cite{Joshi20SA, Wen22PRApp, Fan25LSA, Yan25npj, Zhu25SC}. At present, the research on CVQKD networks for multi-user communication mainly focuses on quantum conferences, where multiple users share the same key simultaneously. Such quantum conference protocols generally utilize the multi-party correlation of CV Greenberger–Horne–Zeilinger (GHZ) entangled states or entanglement swapping to achieve key sharing among multiple users~\cite{Wu16PRA, Zhang18JPA, Otta19CP, Zhao21LPL, Gras22PRXQ, Qin23PR, Zhao25PRA}, which have got certain breakthroughs and progress in theory and experiments. But the quantum conference protocols still cannot realize a CVQKD network with multi-user full connectivity and simultaneous communication. New network schemes and technologies still need to be studied and broken through.

Optical frequency comb possess parallel multi-mode optical field, which can not only be used to boost CVQKD key rates~\cite{Wang19OE} but also be a good source for multi-user quantum communication~\cite{Peng25Math}. In recent years, the multi-mode optical parametric oscillator (OPO) process with a frequency comb structure has begun to attract attention~\cite{Dunl06PRA, Yang13JOSAB, Ros14NP, Chen14PRL, Li19APS, Liu20APS, Shi20PRL, Li24PRApp}. Multiple applications of continuous variable quantum networks based on an entangled frequency comb have been proposed, such as the quantum teleportation network~\cite{Shi23LPR}, the quantum dense coding network~\cite{Liu25SC}, the quantum network coding~\cite{Zhou25PRApp} and so on. In particular, the below-threshold non-degenerate optical parametric amplifier (NOPA) based on the type-$\mathrm{II}$ OPO process can directly generate spatially separated Einstein-Podolsky-Rosen (EPR) entanglement with a frequency comb structure~\cite{Yang13JOSAB, Liu20APS}. Such entangled states can serve as a desired source for the CVQKD network. 

Based on the EPR entangled states with a frequency comb structure generated by the type-$\mathrm{II}$ below-threshold NOPA and combined with the CVQKD protocol with the entangled source in the middle, this paper proposes a multi-user fully connected CVQKD network. By utilizing the characteristic that the polarization states of the signal field and the idler field are perpendicular to each other, the signal comb and the idler comb can be separated easily. With well wavelength management at the central node, all the signal and idler comb teeth can be reasonably allocated and combined, so that all users can share a pair of signal-idler EPR entangled states between each other. Consequently, a star-shaped network structure is able to build between the central node and the users, forming a multi-user fully connected and simultaneously communicable CVQKD network among the users. The technical noise caused by the seed laser and the jitter of the OPO cavity length only affects the low-frequency sidebands with a frequency difference of the order of MHz from the frequency of the seed laser, while the influence on the high-frequency sidebands with a difference of GHz or more can be ignored. The simulation results show that under the conditions of well system loss and noise control, the proposed scheme has the feasibility of deploying a short-distance fully connected CVQKD network. The scheme proposed in this paper provides a solution for establishing a multi-user CVQKD network.

The arrangement of this paper is as follows: In Section \ref{Section-2}, we detail the CVQKD network scheme based on the entangled optical frequency combs. Section \ref{Section-3} presents the security analysis of the proposed scheme. In Section \ref{Section-4}, we show the simulation results of the proposed scheme. Finally, we draw a conclusion in Section \ref{conclusion}.

\section{CVQKD network based on optical frequency comb entangled states}\label{Section-2}

\subsection{Entangled optical frequency comb based on type-II OPO process}\label{Section-2-1}
\begin{figure}[htbp]
\centering
\includegraphics[width=\linewidth]{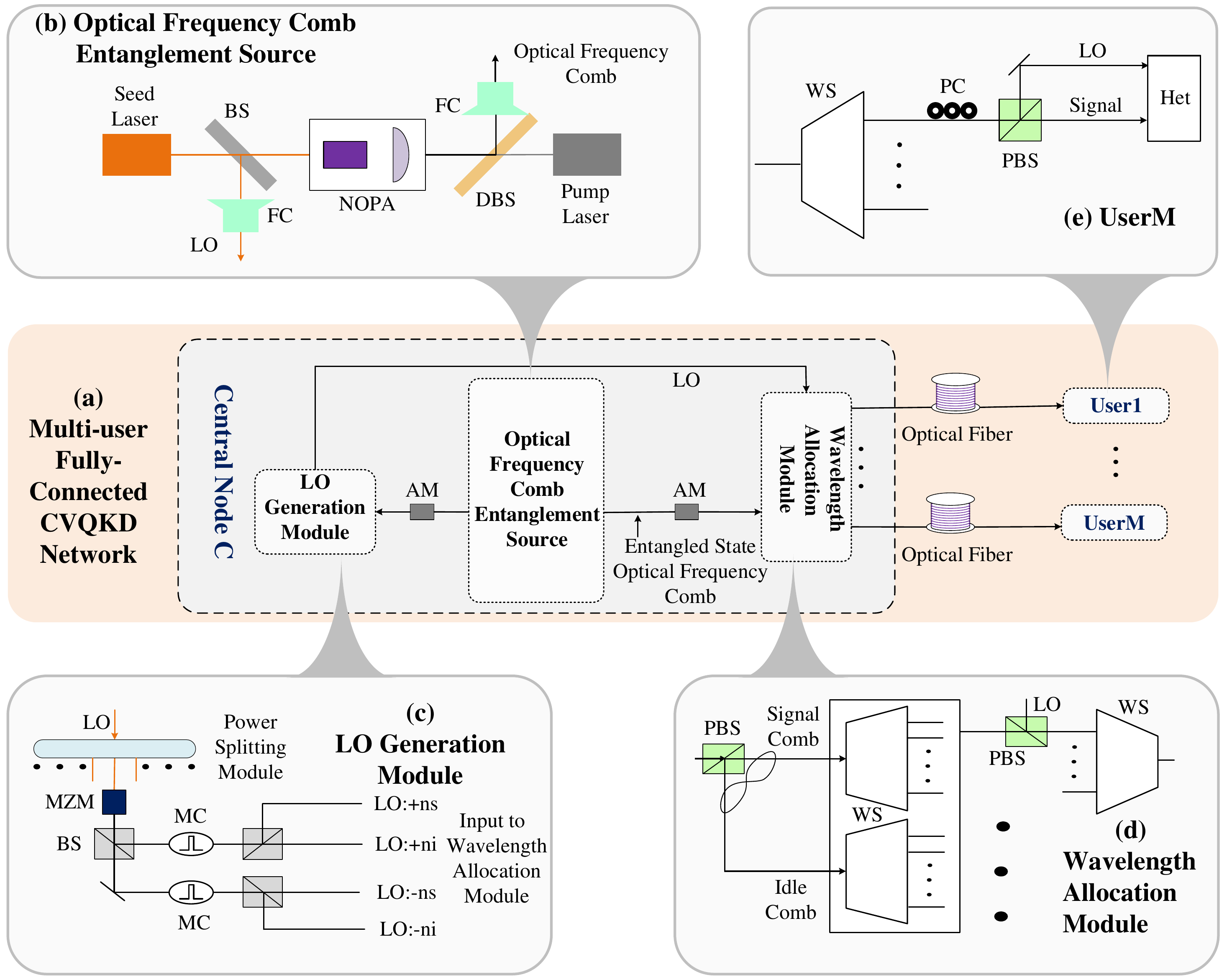}
\caption{Schematic of the optical frequency comb entanglement state based CVQKD networks. $\pm n$s(i) represents the signal (idler) comb with frequency of $\omega_0 \pm n\Omega$. NOPA, nondegenerate optical parametric amplifier; BS, beamsplitter; AM, amplitude modulator; DBS, dichroic beamsplitter; PBS, polarization beamsplitter; PC, polarization controller; MC, mode cleaner; WS, waveshaper; MZM, Mach-Zehnder Modulator; Het, heterodyne detection.}
\label{Fig1}
\end{figure}

Existing studies have shown that the type-$\mathrm{II}$ OPO process can directly generate spatially separated EPR entanglement with a frequency comb structure~\cite{Yang13JOSAB, Liu20APS}, as shown in Fig~\ref{Fig1} (b). The core of the OPO process is the nondegenerate optical parametric amplifier (NOPA), which mainly consists of a cavity formed by a type-II nonlinear crystal and an output coupling mirror. A continuous pump laser ($\omega_ {\mathrm{p}}$) and a seed laser ($\omega_{0} = \omega_{\mathrm{p}}/2$) are simultaneously injected into the NOPA and controlled to have their relative phase equal to $\pi$. When the energy conservation and cavity resonance conditions $\omega_{\mathrm{p}} = \omega_{\mathrm{s}} + \omega_{\mathrm{i}}$ are satisfied, a parametric down-conversion process operating at the deamplification state occurs in the cavity, generating an optical frequency comb with EPR entanglement between the signal and the idler. Among them, the signal frequency comb is $\omega_{\mathrm{s}} = \omega_{0} \mp n\Omega$, and the idler frequency comb is $\omega_{\mathrm{i}} = \omega_{0} \pm n\Omega$, where $\Omega$ is the free spectral range of the OPO, and $n$ is the index of the signal and idler comb tooth pairs. Since the polarization states of the signal and the idler are perpendicular to each other, the signal comb and the idler comb can be spatially separated by a polarizing beam splitter (PBS). Note that the total bandwidth of the optical frequency comb is determined by the phase-matching bandwidth of the NOPA. For a nonlinear medium with a length of $L_{\mathrm{cry}}$, the down-conversion bandwidth can be estimated as $10c/L_{\mathrm{cry}}$. For an NOPA with a cavity length of $L_{\mathrm{cav}}$, the free spectral range can be calculated as $\Omega = c/2L_{\mathrm{cav}}$~\cite{Raym05PRA}. Therefore, the lengths of the nonlinear medium and the cavity can be designed according to the actual requirements of the bandwidth and the free spectral range.

\subsection{Multi-user fully connected CVQKD network}
By utilizing the entangled optical frequency comb generated based on the type-II OPO process and combining it with the CVQKD scheme with the entangled source in the middle~\cite{Weed13PRA, Guo17PRA}, a multi-user fully connected CVQKD network can be established, as shown in Fig.~\ref{Eq-1} (a). In this network, a star-shaped network structure is formed between the central node C and the users, and all users are connected in a fully connected network. Specifically, for the central node C, the entangled source of optical frequency comb couples the comb generated by the OPO process into an optical fiber through an optical fiber coupler. Then, an amplitude modulator (AM) is used for chopping to generate optical frequency comb pulses, which are then sent to the wavelength allocation module (Fig.~\ref{Fig1} (d)). In the wavelength allocation module, the PBS first separates the entangled signal comb and idler comb pulses from the optical frequency comb pulses. Then, the two combs are sent to two waveshapers respectively for comb-tooth separation and combination, so as to distribute them to different users to form a multi-user fully connected CVQKD network.

On the other hand, before being sent to the NOPA, the seed laser is split into two parts by an unbalanced beam splitter. The part with lower intensity is sent to the NOPA, and the other part is used as the local oscillator (LO). First, it is coupled into an optical fiber through an optical fiber coupler (FC), then chopped by an AM to form optical pulses. The generated pulses are then sent to the LO generation module (Fig.~\ref{Fig1}v(c)) to generate the LO required by the system. In the LO generation module, the optical pulses are evenly divided into $N(N-1)/2$ parts by an optical power splitting module (which can be realized by a combination of beam splitters). Each part is first subjected to carrier-suppressed double-sideband modulation by a Mach-Zehnder modulator (MZM) to obtain double-sideband pulses with frequencies of $\omega_0 \pm k\Omega (k = 1,2,...)$. Then, each pulse is split into two beams by a balanced beam splitter (BS). Each beam passes through a mode cleaner (MC) to obtain positive or negative frequency sidebands. Each sideband pulse is further split into two parts by a balanced beam splitter and used as the LO for the signal and idler at the comb-tooth frequency of $\omega_0 + k\Omega (k = 1,2,...)$ or $\omega_0 - k\Omega (k = 1,2,...)$. Note that there are signal and idler comb teeth with mutually orthogonal polarization states at each comb-tooth frequency. Through the above process, we obtain the EPR entangled pairs and their corresponding LOs for CVQKD.

\begin{figure}[htbp]
\centering\includegraphics[width=0.55\linewidth]{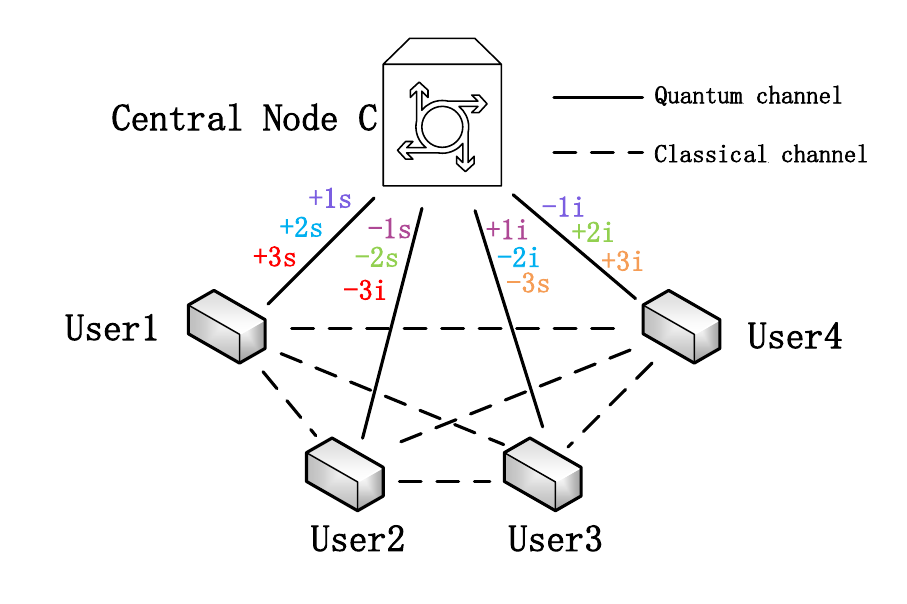}
\caption{Schematic of the optical frequency comb entanglement state based 4 users CVQKD networks. $\pm n$s(i) represents the signal (idler) comb with frequency of $\omega_0 \pm n\Omega$, where $n = 1,2,3$.}
\label{Fig2}
\end{figure}

Suppose there are $N$ users. We divide the total signal and idle frequency combs into $N$ groups. Each signal comb or idler comb is polarization-division multiplexed in time division with the corresponding LO through a PBS, and then wavelength-division multiplexed through $N$ waveshapers. After that, the multiplexed signals are sent to $N$ users respectively. After receiving the signals, each receiving user (Fig. ~\ref{Fig1} (e)) first performs wavelength-division demultiplexing through an waveshapers to obtain $N - 1$ channels of signals. For each frequency signal, the receiver uses a polarization controller (PC) to adjust the polarization state of the optical signal, then separates the quantum signal and the LO through a PBS. After that, the polarization state of the LO is rotated by 90$^\circ$ and then sent together with the quantum signal to the detection module for coherent detection. Finally, the users generate the final key through post-data processing. Fig.~\ref{Fig2} shows a schematic diagram of a fully-connected CVQKD network for 4 users.

\section{Security analysis}\label{Section-3}
Next, we will take the security analysis between two users, $\mathrm{U}_1$ (referred to as A) and $\mathrm{U}_2$ (referred to as B), as an example to illustrate the security of the CVQKD network. The security between other users is similar for the star-shaped CVQKD network. For central node C and users A and B, node C distributes the two modes of the EPR entangled pair to users A and B, respectively. This process is equivalent to a CVQKD scheme with an entanglement source in the middle. Therefore, our security analysis can be discussed based on the conclusions in Refs.~\cite{Weed13PRA, Guo17PRA}.

The signal and idler modes with EPR entanglement form a two-mode squeezed state (TMSS). To analyze the security of the system, we first need to obtain the covariance matrix of the TMSS generated by the OPO process. For this purpose, we need to obtain the expressions of the quadratures of the signal and idler modes. Under the conditions of perfect phase matching and no cavity detuning, the quantum Langevin equations of motion for the intracavity fields in the OPO process are given as~\cite{Yang13JOSAB, Liu20APS}
\begin{eqnarray}\label{Eq-1}
\begin{aligned}
\dot{\hat{a}}_{\mathrm{i}}(t)= & -k_{1} \hat{a}_{\mathrm{i}}(t)-\chi\hat{a}_{\mathrm{s}}^{\dagger}(t) +\sqrt{2 k} \hat{b}_{\mathrm{i}}^{\mathrm{in}}(t)+\sqrt{2 \gamma} \hat{c}_{\mathrm{i}}(t), \\
\dot{\hat{a}}_{\mathrm{s}}(t)= & -k_{1} \hat{a}_{\mathrm{s}}(t)-\chi\hat{a}_{\mathrm{i}}^{\dagger}(t) + \sqrt{2 k} \hat{b}_{\mathrm{s}}^{\mathrm{in}}(t)+\sqrt{2 \gamma} \hat{c}_{\mathrm{s}}(t),
\end{aligned}
\end{eqnarray}
where $\hat{a}_{\mathrm{i}}$ is the annihilation operator of the down-conversion field. The subscript $j = \mathrm{i}, \mathrm{s}$ represents the idler field and the signal field. $\hat{b}_{j}$ represents the input signal and idler fields, and $\hat{c}_{j}$ represents the vacuum noise introduced by the intracavity loss. $\chi$ is the single-pass parametric amplitude gain, which is proportional to the second-order nonlinear coupling coefficient and the amplitude of the pump light. $k$ and $\gamma$ represent the transmission loss rate of the down-conversion field at the output coupling mirror and other additional intracavity loss rates, respectively. $k_1 = k + \gamma$ is the total loss rate. In the steady-state case, after the intracavity field makes a single cavity round trip, the operator equation can be expressed as \cite{Dunl06PRA, Yang13JOSAB}
\begin{eqnarray}\label{Eq-2}
\begin{aligned}
\hat{a}_{\mathrm{i}}(t + \tau)= & [-\chi\tau\hat{a}_{\mathrm{s}}^{\dagger}(t) + (1 - k_{1}\tau) \hat{a}_{\mathrm{i}}(t) + \sqrt{2 k}\tau \hat{b}_{\mathrm{i}}^{\mathrm{in}}(t)+\sqrt{2 \gamma} \tau \hat{c}_{\mathrm{i}}(t)]e^{i\Delta\tau} \\
\approx & [-\chi\tau\hat{a}_{\mathrm{s}}^{\dagger}(t) + (1 - k_{1}\tau) \hat{a}_{\mathrm{i}}(t) + \sqrt{2 k}\tau \hat{b}_{\mathrm{i}}^{\mathrm{in}}(t)+\sqrt{2 \gamma} \tau \hat{c}_{\mathrm{i}}(t)]e^{i\bar{\Delta}\tau}(1 + i\tau\delta\Delta(t)), \\
\hat{a}_{\mathrm{s}}(t + \tau)= & [-\chi\tau\hat{a}_{\mathrm{i}}^{\dagger}(t) + (1 -k_{1}\tau) \hat{a}_{\mathrm{s}}(t) + \sqrt{2 k}\tau \hat{b}_{\mathrm{s}}^{\mathrm{in}}(t)+\sqrt{2 \gamma} \tau \hat{c}_{\mathrm{s}}(t)]e^{i\Delta\tau} \\
\approx & [-\chi\tau\hat{a}_{\mathrm{i}}^{\dagger}(t) + (1 -k_{1}\tau) \hat{a}_{\mathrm{s}}(t) + \sqrt{2 k}\tau \hat{b}_{\mathrm{s}}^{\mathrm{in}}(t)+\sqrt{2 \gamma} \tau \hat{c}_{\mathrm{s}}(t)]e^{i\bar{\Delta}\tau}(1 + i\tau\delta\Delta(t)),
\end{aligned}
\end{eqnarray}
where $\tau = 1/\Omega$ is the time for the light field to travel back and forth in the cavity for one round, $\Delta = \bar{\Delta} + \delta\Delta(t)$ is the cavity detuning, $\bar{\Delta}$ is the average cavity detuning, $\delta\Delta(t)$ is the time-varying fluctuations of cavity detuning, which are caused by the actual jitter of the OPO cavity length. In the above equation, we assume that $\delta\Delta(t) \ll \bar{\Delta}$ and make a first-order approximation for $\delta\Delta(t)$~\cite{Dunl06PRA}. Decompose the operator of the field into the sum of the mean value and the quantum fluctuation, that is, $\hat{a} = \langle\hat{a}\rangle + \delta\hat{a}$. Since $\langle\hat{a}(t + \tau)\rangle = \langle\hat{a}(t)\rangle$ in the steady state, the amplitude of the intracavity field can be obtained from equation (\ref{Eq-2})
\begin{eqnarray}\label{Eq-3}
\langle\hat{a}_{\mathrm{s}({\mathrm{i}})}\rangle = \frac{1}{k_1^2 - \chi^2}[-\sqrt{2k}\chi\langle b_{\mathrm{i}({\mathrm{s}})}^{in\dagger}\rangle + \sqrt{2k}k_1\langle b_{\mathrm{s}({\mathrm{i}})}^{in}\rangle].
\end{eqnarray}
Here we have assumed that $\bar{\Delta} = 0$.

According to equations (\ref{Eq-2}) - (\ref{Eq-3}) and the cavity field input-output relation $\hat{a}^{out} = \sqrt{2 k}a - b^{in}$, combined with the frequency-domain analysis method in Ref.\cite{Yang13JOSAB}, we can obtain the expressions of the quadratures of the down-converted signal and the idler field
\begin{eqnarray}
\hat{X}_{\mathrm{s}(\mathrm{i})}^{out}(\omega) &=& \langle\hat{X}_{\mathrm{s}(\mathrm{i})}^{out}(\omega)\rangle + \frac{-\chi[2k\delta\hat{X}_{\mathrm{i}(\mathrm{s})}^{b^{in}}(\omega) + \sqrt{4k\gamma}\delta\hat{X}_{\mathrm{i}(\mathrm{s})}^{c}(\omega)]}{(\frac{e^{i\omega\tau} - 1}{\tau} + k_1)^2 - \chi^2}  \nonumber \\
& & + \frac{[k^2 + \chi^2 - (\frac{e^{i\omega\tau} - 1}{\tau} + \gamma)^2]\delta\hat{X}_{\mathrm{s}(\mathrm{i})}^{b^{in}}(\omega)}{(\frac{e^{i\omega\tau} - 1}{\tau} + k_1)^2 - \chi^2} \nonumber \\
& & + \frac{\sqrt{4k\gamma}(\frac{e^{i\omega\tau} - 1}{\tau} + k_1)\delta\hat{X}_{\mathrm{s}(\mathrm{i})}^{c}(\omega)}{(\frac{e^{i\omega\tau} - 1}{\tau} + k_1)^2 - \chi^2} + \frac{\Xi_{\mathrm{s}(\mathrm{i})}^{x}}{k_1^2 - \chi^2}\delta\Delta(\omega), \label{Eq-4} \\
\hat{P}_{\mathrm{s}(\mathrm{i})}^{out}(\omega) &=& \langle\hat{P}_{\mathrm{s}(\mathrm{i})}^{out}(\omega)\rangle + \frac{\chi[2k\delta\hat{P}_{\mathrm{i}(\mathrm{s})}^{b^{in}}(\omega) + \sqrt{4k\gamma}\delta\hat{P}_{\mathrm{i}(\mathrm{s})}^{c}(\omega)]}{(\frac{e^{i\omega\tau} - 1}{\tau} + k_1)^2 - \chi^2} \nonumber \\
& & + \frac{[k^2 + \chi^2 - (\frac{e^{i\omega\tau} - 1}{\tau} + \gamma)^2]\delta\hat{P}_{\mathrm{s}(\mathrm{i})}^{b^{in}}(\omega)}{(\frac{e^{i\omega\tau} - 1}{\tau} + k_1)^2 - \chi^2} \nonumber \\
& & + \frac{\sqrt{4k\gamma}(\frac{e^{i\omega\tau} - 1}{\tau} + k_1)\delta\hat{P}_{\mathrm{s}(\mathrm{i})}^{c}(\omega)}{(\frac{e^{i\omega\tau} - 1}{\tau} + k_1)^2 - \chi^2} + \frac{\Xi_{\mathrm{s}(\mathrm{i})}^{p}}{k_1^2 - \chi^2}\delta\Delta(\omega), \label{Eq-5}
\end{eqnarray}
where
\begin{eqnarray}
\langle\hat{X}_{\mathrm{s}(\mathrm{i})}^{out}(\omega)\rangle &=& \frac{1}{k_1^2 - \chi^2}[-2k\chi\langle \hat{X}_{\mathrm{i}(\mathrm{s})}^{b^{in}}(\omega)\rangle + (k^2 + \chi^2 - \gamma^2)\langle\hat{X}_{\mathrm{s}(\mathrm{i})}^{b^{in}}(\omega)\rangle], \\
\langle\hat{P}_{\mathrm{s}(\mathrm{i})}^{out}(\omega)\rangle &=& \frac{1}{k_1^2 - \chi^2}[-2k\chi\langle \hat{P}_{\mathrm{i}(\mathrm{s})}^{b^{in}}(\omega)\rangle + (k^2 + \chi^2 - \gamma^2)\langle\hat{P}_{\mathrm{s}(\mathrm{i})}^{b^{in}}(\omega)\rangle],\\
\Xi_{\mathrm{s}(\mathrm{i})}^{x} &=& - 2k\chi\tau(e^{i\omega\tau} - 1)\langle\hat{P}_{\mathrm{i}(\mathrm{s})}^{b^{in}}(\omega)\rangle - [2k\tau^2(k_1^2 - \chi^2) \nonumber \\
& & + 2kk_1\tau(e^{i\omega\tau} - 1)]\langle\hat{P}_{\mathrm{s}(\mathrm{i})}^{b^{in}}(\omega)\rangle, \\
\Xi_{\mathrm{s}(\mathrm{i})}^{p} &=&  2k\chi\tau(e^{i\omega\tau} - 1)\langle\hat{X}_{\mathrm{i}(\mathrm{s})}^{b^{in}}(\omega)\rangle + [2k\tau^2(k_1^2 - \chi^2) \nonumber \\
& &  + 2kk_1\tau(e^{i\omega\tau} - 1)]\langle\hat{X}_{\mathrm{s}(\mathrm{i})}^{b^{in}}(\omega)\rangle,
\end{eqnarray}
$\langle\hat{X}_{j}^{b^{in}}(\omega)\rangle = \langle\hat{b}_{j}^{in \dagger}(\omega)\rangle + \langle\hat{b}_{j}^{in}(\omega)\rangle$ and $\langle\hat{P}_{j}^{b^{in}}(\omega)\rangle = i[\langle\hat{b}_{j}^{in \dagger}(\omega)\rangle - \langle\hat{b}_{j}^{in}(\omega)\rangle]$ ($j = \mathrm{s},\mathrm{i}$) are the mean values of the amplitude and phase quadratures of the input field, respectively. $\delta\hat{X}_{j}^{b^{in}}(\omega) = \delta\hat{b}_{j}^{in \dagger}(\omega) + \delta\hat{b}_{j}^{in}(\omega)$ and $\delta\hat{P}_{j}^{b^{in}}(\omega) = i[\delta\hat{b}_{j}^{in \dagger}(\omega) - \delta\hat{b}_{j}^{in}(\omega)]$ are quantum fluctuations of the amplitude and phase quadratures of the input field, respectively. $\delta\hat{X}_{j}^{c}(\omega) = \delta\hat{c}_{j}^{\dagger}(\omega) + \delta\hat{c}_{j}(\omega)$ and $\delta\hat{P}_{j}^{c}(\omega) = i[\delta\hat{c}_{j}^{\dagger}(\omega) - \delta\hat{c}_{j}(\omega)]$ is the quantum fluctuations of the field's amplitude and phase quadratures introduced by intracavity losses, respectively. $\omega$ is the analysis frequency.

Based on equations (\ref{Eq-4}) - (\ref{Eq-5}), we can directly calculate the covariance matrix of each pair of TMSS output from the OPO process. For each pair of generated TMSS, since the cavity is resonant with the OPO seed, we have $\omega\tau = \pm 2\pi n$. In this case, each matrix element of the covariance matrix can be calculated as
\begin{align}
\mathrm{Var}(\hat{X}_{\mathrm{s}(\mathrm{i})}^{out}) &= \frac{4kk_1\chi^2 + (k^2 + \chi^2 - \gamma^2)^2 + 4k\gamma k_1^2}{(k_1^2 - \chi^2)^2} + \tilde{N}_{\mathrm{s}(\mathrm{i})}^{in}(\omega) + \tilde{N}_{\mathrm{s}(\mathrm{i})}^{\delta\Delta,x}(\omega) \equiv V_0^X,  \label{Eq-7} \\
\mathrm{Var}(\hat{P}_{\mathrm{s}(\mathrm{i})}^{out}) &= \frac{4kk_1\chi^2 + (k^2 + \chi^2 - \gamma^2)^2 + 4k\gamma k_1^2}{(k_1^2 - \chi^2)^2} + \tilde{N}_{\mathrm{s}(\mathrm{i})}^{in}(\omega) + \tilde{N}_{\mathrm{s}(\mathrm{i})}^{\delta\Delta,p}(\omega) \equiv V_0^P,  \label{Eq-8} \\
\mathrm{Cov}(\hat{X}_{\mathrm{s}}^{out}\hat{X}_{\mathrm{i}}^{out}) &= \frac{-4k\chi(k^2 + \chi^2 - \gamma^2) - 8kk_1\gamma\chi}{(k_1^2 - \chi^2)^2} + \tilde{C}_N^{in,x} \equiv C_{0}^X, \label{Eq-9}  \\ \mathrm{Cov}(\hat{P}_{\mathrm{s}}^{out}\hat{P}_{\mathrm{i}}^{out}) &= -\frac{-4k\chi(k^2 + \chi^2 - \gamma^2) - 8kk_1\gamma\chi}{(k_1^2 - \chi^2)^2} + \tilde{C}_N^{in,p} \equiv C_{0}^P, \label{Eq-10}  \\
\mathrm{Cov}(\hat{X}_{\mathrm{s}}^{out}\hat{P}_{\mathrm{i}}^{out}) &= \tilde{C}_{\delta\Delta}^{xp}, \ \
\mathrm{Cov}(\hat{P}_{\mathrm{s}}^{out}\hat{X}_{\mathrm{i}}^{out}) = \tilde{C}_{\delta\Delta}^{px},\label{Eq-11}
\end{align}
where $\mathrm{Var}(\digamma) = \langle\digamma\digamma^{\dagger}\rangle - \langle\digamma\rangle\langle\digamma^{\dagger}\rangle$, $\mathrm{Cov}(\mu\nu) = \langle\mu\nu^{\dagger}\rangle - \langle\mu\rangle\langle\nu^{\dagger}\rangle$, $\langle|\delta\hat{X}_{j}^{c}(\omega)|^2\rangle = \langle|\delta\hat{P} _{j}^{c}(\omega)|^2\rangle = 1$ (shot noise units),
\begin{align}
\tilde{N}_{\mathrm{s}(\mathrm{i})}^{in}(\omega) &= \frac{4k^2\chi^2N_{\mathrm{i}(\mathrm{s})}^{b^{in}}(\omega) + (k^2 + \chi^2 - \gamma^2)N_{\mathrm{s}(\mathrm{i})}^{b^{in}}(\omega)}{(k_1^2 - \chi^2)^2}, \\
\tilde{N}_{\mathrm{s}(\mathrm{i})}^{\delta\Delta,x}(\omega) &= 4k^2\tau^4\langle\hat{P}_{\mathrm{s}(\mathrm{i})}^{b^{in}}(\omega)\rangle^2\sigma_{\delta\Delta}^2,
\tilde{N}_{\mathrm{s}(\mathrm{i})}^{\delta\Delta,p}(\omega) = 4k^2\tau^4\langle\hat{X}_{\mathrm{s}(\mathrm{i})}^{b^{in}}(\omega)\rangle^2\sigma_{\delta\Delta}^2,\\
\tilde{C}_N^{in,x} &= \frac{-2k\chi(k^2 + \chi^2 - \gamma^2)[N_{\mathrm{i}}^{b^{in}}(\omega) + N_{\mathrm{s}}^{b^{in}}(\omega)]}{(k_1^2 - \chi^2)^2} - 4k^2\tau^4\langle\hat{P}_{\mathrm{s}}^{b^{in}}(\omega)\rangle \langle\hat{P}_{\mathrm{i}}^{b^{in}}(\omega)\rangle\sigma_{\delta\Delta}^2,\\
\tilde{C}_N^{in,p} &= \frac{-2k\chi(k^2 + \chi^2 - \gamma^2)[N_{\mathrm{i}}^{b^{in}}(\omega) + N_{\mathrm{s}}^{b^{in}}(\omega)]}{(k_1^2 - \chi^2)^2} - 4k^2\tau^4\langle\hat{X}_{\mathrm{s}}^{b^{in}}(\omega)\rangle \langle\hat{X}_{\mathrm{i}}^{b^{in}}(\omega)\rangle\sigma_{\delta\Delta}^2,\\
\tilde{C}_{\delta\Delta}^{xp} &= -4k^2\tau^4\langle\hat{P}_{\mathrm{s}}^{b^{in}}(\omega)\rangle \langle\hat{X}_{\mathrm{i}}^{b^{in}}(\omega)\rangle\sigma_{\delta\Delta}^2,
\tilde{C}_{\delta\Delta}^{px} = -4k^2\tau^4\langle\hat{P}_{\mathrm{i}}^{b^{in}}(\omega)\rangle \langle\hat{X}_{\mathrm{s}}^{b^{in}}(\omega)\rangle\sigma_{\delta\Delta}^2,
\end{align}
$\sigma_{\delta\Delta}^2 = \langle|\delta\Delta(\omega)|^2\rangle$ is the variance of $\delta\Delta(\omega)$. $N_{\mathrm{i}(\mathrm{s})}^{b^{in}}(\omega)$ represents the noise of the input signal and idler fields above the quantum fluctuations, and it is assumed that the noises of the two quadratures are equal, i.e. $\langle|\delta\hat{X} _{j}^{b^{in}}(\omega)|^2\rangle = \langle|\delta\hat{P} _{j}^{b^{in}}(\omega)|^2\rangle = 1 + N_j^{b^{in}}(\omega)$. Here, $\tilde{N}_{\mathrm{s}(\mathrm{i})}^{in}(\omega)$ represents the noise introduced by the extra optical noise of the seed laser. $\tilde{C}_N^{in,x/p}$, $\tilde{C}_{\delta\Delta}^{xp}$, and $\tilde{C}_{\delta\Delta}^{px}$ are the covariance noises introduced by the extra optical noise of the seed laser. $\tilde{N}_{\mathrm{s}(\mathrm{i})}^{\delta\Delta,x}(\omega)$ and $\tilde{N} _{\mathrm{s}(\mathrm{i})}^{\delta\Delta,p}(\omega)$ are the noises introduced by the OPO cavity length jitter.

\begin{figure}[t]
\centering
\subfigure[]{
\includegraphics[width=0.47\linewidth]{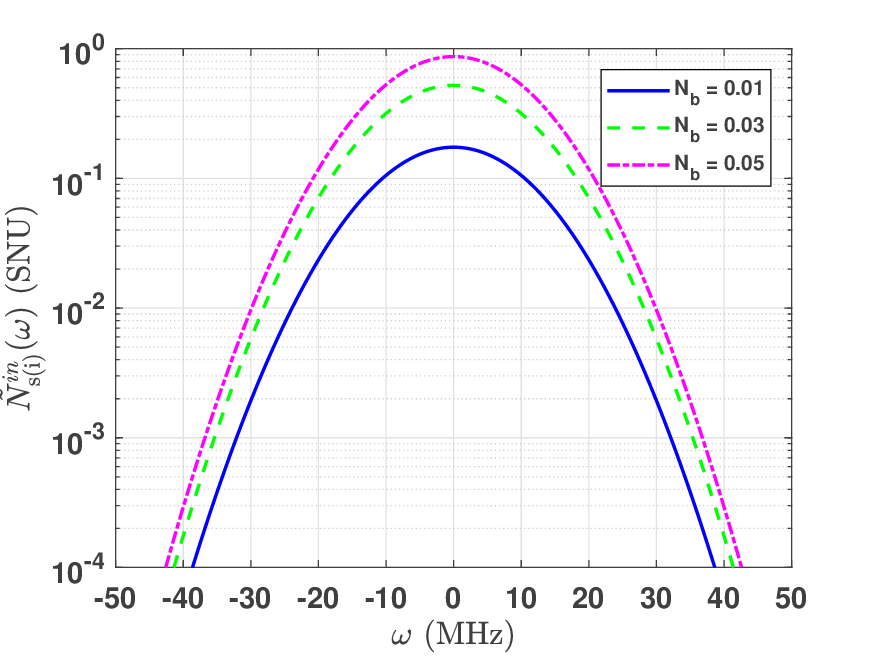}
\label{fig:DGD-L-a}
}
\subfigure[]{
\centering
\includegraphics[width=0.47\linewidth]{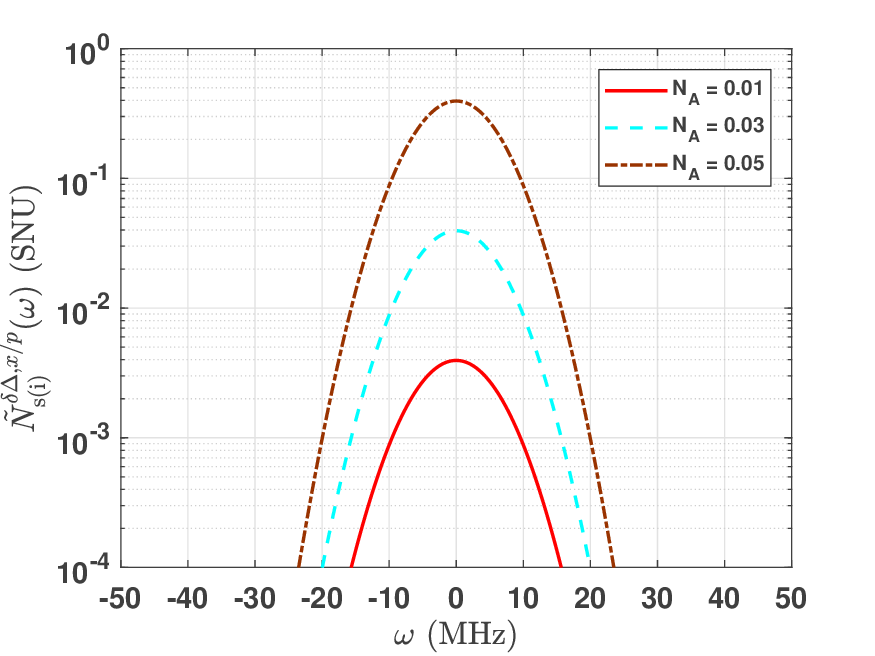}
\label{fig:DGD-L-b}
}
\subfigure[]{
\centering
\includegraphics[width=0.47\linewidth]{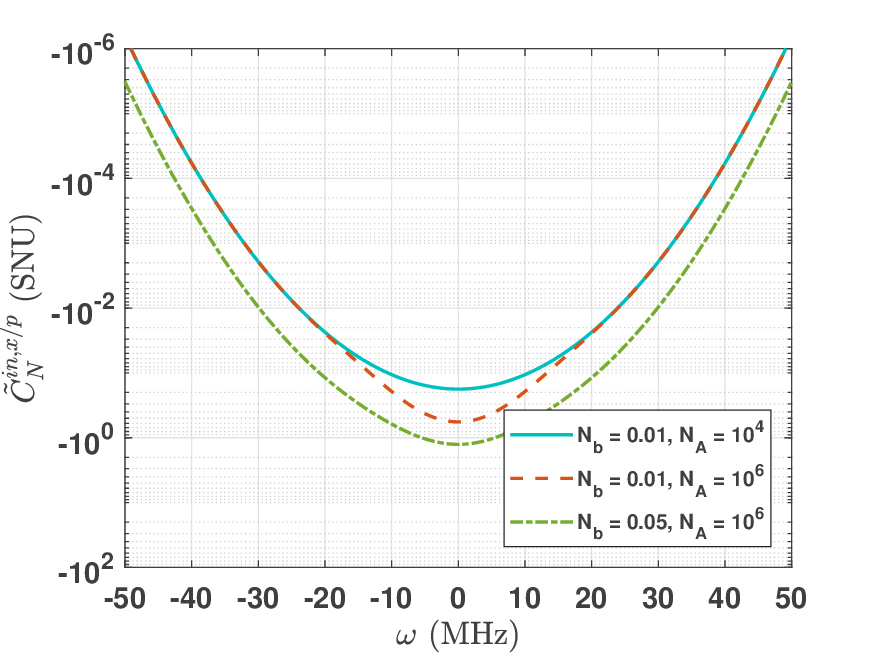}
\label{fig:DGD-L-b}
}
\subfigure[]{
\centering
\includegraphics[width=0.47\linewidth]{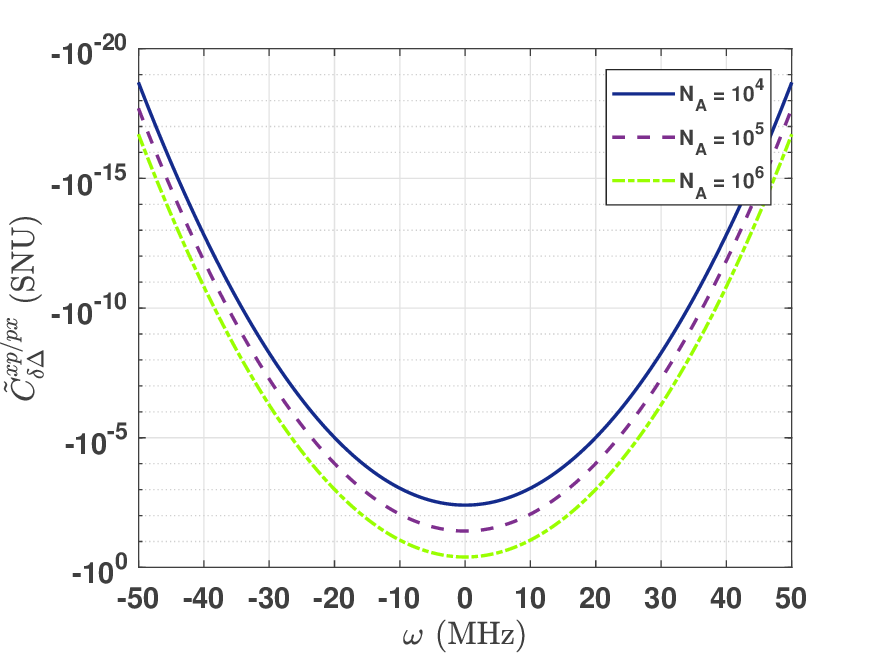}
\label{fig:DGD-L-b}
}
\caption{The noise contributed from the extra optical noise carried by the seed and fluctuations of the length of the OPO cavity as a function of the frequency. (a) The noise introduced by the seed optical noise to the generated TMSS. (b) The noise introduced by fluctuations of the length of the OPO cavity to the generated TMSS. (c) The covariance noise $\tilde{C} _N^{in,x/p}$ introduced by the seed optical noise to the generated TMSS. (d) The covariance noise $\tilde{C} _{\delta\Delta}^{xp/px}$ introduced by the seed optical noise to the generated TMSS.}
\label{fig:DGD-L}
\end{figure}

In the above discussion, we have considered two main technical noise sources, namely the excess optical noise of the seed laser (the part above the shot noise) and the noise introduced by the OPO cavity length jitter \cite{Dunl06PRA}. For the excess noise of the seed laser, since the linewidth of the seed laser can generally be limited within the MHz range, the noise spectrum of this noise can also be assumed to be limited within MHz. If $\Omega$ is much larger than the MHz level, the influence of this noise on the optical frequency comb will be limited to the signal and idler combs at the frequency $\omega_0$, and the influence on the sideband combs with larger frequencies can be ignored. For the noise introduced by the OPO cavity length jitter, with well cavity length locking technique (for example, the Pound-Drever-Hall frequency stabilization technique~\cite{Liu20APS}), the frequency of the cavity length jitter is generally not too high (within the MHz level), and the detuning frequency caused by the jitter can also be controlled within a small range, making $\delta\Delta(t)\tau \ll 2\pi$. In addition, this noise is also related to the amplitudes of the signal and idler input fields. In practice, there is only an injected field with a frequency of $\omega_0$. Therefore, the noise introduced by the cavity length jitter mainly affects the signal and idler combs whose frequency difference from $\omega_0$ is within the MHz level. So if $\Omega$ is much larger than the MHz level, the influence of the noise introduced by the cavity length jitter will also be limited to the signal and idler combs near the frequency $\omega_0$, and has almost no influence on the high-frequency sideband combs. We show the variation of the influence of these two noise sources with respect to $\omega$ in Fig.~\ref{fig:DGD-L}, where we assume that $N_{\mathrm{s}(\mathrm{i})}^{b^{in}}(\omega) = N_be^{-\omega^2/(4\pi\times10^7)}$, $\sigma_{\delta\Delta} = 2\pi\times10^7e^{-\omega^2/(4\pi\times10^7)}$, $\langle\hat{X}_{\mathrm{s}(\mathrm{i})}^{b^{in}}(\omega)\rangle^2 = \langle\hat{P}_{\mathrm{s}(\mathrm{i})}^{b^{in}}(\omega)\rangle^2 = N_Ae^{-\omega^2/(4\pi\times10^7)}$, $\Omega = 10$ GHz, $k\tau = 0.05$\cite{Liu20APS}, $k_1\tau = 0.052$, $\chi = 0.75k$. From the results in Fig.~\ref{fig:DGD-L}, it can be seen that the influence of these two noise sources has a large impact only in the region where $|\omega|$ is small (within the MHz level), and the influence on the high - frequency sidebands (GHz level or above) can be ignored.

Based on the above discussions and results, we can now write the covariance matrix of the high - frequency sideband EPR entangled pairs generated by the OPO process. Denote the TMSS composed of a pair of signal and idler fields generated by the OPO process as $\rho_{A_0B_0}$, where $A_0$ and $B_0$ represent the two modes of the TMSS. Then its covariance matrix can be expressed as
\begin{eqnarray}\label{Eq-9}
\Upsilon_{A_0B_0} = \left(
                      \begin{array}{cc}
                        V_0\mathbb{I} & C_{0}\sigma_z \\
                        C_{0}\sigma_z & V_0\mathbb{I} \\
                      \end{array}
                    \right),
\end{eqnarray}
where $V_0 = \frac{4kk_1\chi^2 + (k^2 + \chi^2 - \gamma^2)^2 + 4k\gamma k_1^2}{(k_1^2 - \chi^2)^2}$, $C_0 = \frac{-4k\chi(k^2 + \chi^2 - \gamma^2) - 8kk_1\gamma\chi}{(k_1^2 - \chi^2)^2}$, $\mathbb{I} = \mathrm{diag}(1,1)$, $\sigma_z = \mathrm{diag}(1,-1)$. Furthermore, if the additional loss rate of the inner cavity of the NOPA is 0, i.e., $\gamma = 0$, we find that $C_0 = \sqrt{V_0^2 - 1}$. That is to say, the generated TMSS is a two-mode squeezed vacuum state.

Before the state $\rho_{A_0B_0}$ is emitted from the central node C, it needs to pass through a series of optical components (including fiber couplers, amplitude modulators, polarization beam splitters, waveshaper, polarization controllers, etc.). Assume that the total transmittances of modes $A_0$ and $B_0$ passing through these optical components are $\eta_1$ and $\eta_2$ respectively. Let $\rho_{A_1B_1}$ represent the TMSS when it is emitted from the central node C. Then, the covariance matrix of this state at this time can be expressed as
\begin{eqnarray}\label{Eq-10}
\Upsilon_{A_1B_1} = \left(
                      \begin{array}{cc}
                        \eta_1V_0 + 1 - \eta_1\mathbb{I} & \sqrt{\eta_1\eta_2}C_{0}\sigma_z \\
                        \sqrt{\eta_1\eta_2}C_{0}\sigma_z & \eta_2V_0 + 1 - \eta_2 \mathbb{I} \\
                      \end{array}
                    \right) = \left(
                                \begin{array}{cc}
                                  V_1\mathbb{I} & C_{12}\sigma_z \\
                                  C_{12}\sigma_z & V_2\mathbb{I} \\
                                \end{array}
                              \right),
\end{eqnarray}
where $V_1 = \eta_1V_0 + 1 - \eta_1$, $V_2 = \eta_2V_0 + 1 - \eta_2$, $C_{12} = \sqrt{\eta_1\eta_2}C_{0}$. After the state $\rho_{A_1B_1}$ is transmitted through the channel, coherent detection is performed at the receiver, obtaining the state $\rho_{A_2B_2}$. Its covariance matrix becomes
\begin{eqnarray}\label{Eq-11}
\Upsilon_{A_2B_2} = \left(
                                \begin{array}{cc}
                                  a\mathbb{I} & c\sigma_z \\
                                  c\sigma_z & b\mathbb{I} \\
                                \end{array}
                              \right),
\end{eqnarray}
where $a = \eta_D\eta T_1V_1 + \eta_D\eta(1 - T_1)W_1 + 1 - \eta_D\eta + \upsilon_{el}$, $b = \eta_D\eta T_2V_2 + \eta_D\eta(1 - T_2)W_2 + 1 - \eta_D\eta + \upsilon_{el}$, $c = \eta\eta_D\sqrt{T_1T_2}C_{12}$. $T_1$ and $T_2$ are the transmittances of modes $A_1$ and $B_1$ on the channel respectively. $\eta$ and $\upsilon_{el}$ are the efficiency and electrical noise of the detector, respectively. Here, we assumed that the detectors of the two users have the same values. $W_1 = 1 + T_1\varepsilon_1/(1 - T_1)$, $W_2 = 1 + T_2\varepsilon_2/(1 - T_2)$, where $\varepsilon_1$ and $\varepsilon_2$ are the excess noises introduced by the two channels, and $\eta_D$ is the insertion loss of the waveshapers at the two user ends.

Now we can evaluate the security of CVQKD between two users. According to the analysis and conclusions in the literature \cite{Weed13PRA, Guo17PRA}, for the CVQKD scheme with an entangled source in the middle, the direct reconciliation protocol (with A as the reference point), where user A performs heterodyne detection and user B performs homodyne detection, has better performance. Next, we will calculate the key rate based on this scenario. The key rate between A and B can be calculated as
\begin{eqnarray}
  K = \beta I_{AB} - \chi_{AE},
\end{eqnarray}
where $\beta$ is the reconciliation efficiency, $I_{AB}$ is the mutual information between A and B, $\chi_{AE}$ is the Holevo bound of mutual information between A and Eve. The mutual information between A and B can be written as
\begin{eqnarray}
  I_{AB} = \frac{1}{2}\log_2\left(\frac{a + 1}{a + 1 - c^2/b}\right).
\end{eqnarray}
For $\chi_{AE}$,
\begin{eqnarray}
  \chi_{AE} = G(\frac{\nu_1 - 1}{2}) + G(\frac{\nu_2 - 1}{2}) - G(\frac{\nu_3 - 1}{2}) - G(\frac{\nu_4 - 1}{2}).
\end{eqnarray}
The symplectic eigenvalues $\nu_{1,2}$ can be computed as
\begin{eqnarray}
  \nu_{1,2}^2 = \frac{1}{2}[\Delta \pm \sqrt{\Delta^2 - 4D^2}],
\end{eqnarray}
where $\Delta = a^2 + b^2 - c^2$, $D = ab - c^2$. For symplectic eigenvalues $\nu_{3,4}$,
\begin{eqnarray}
  \nu_{3,4}^2 = \frac{1}{2}[\mathbb{A} \pm \sqrt{\mathbb{A}^2 - 4\mathbb{B}}],
\end{eqnarray}
where $\mathbb{A} = (a + bD + \Delta)/(a + 1)$, $\mathbb{B} = D(b + D)/(a + 1)$.

\section{Simulation results and discussion}\label{Section-4}
Based on the previous security analysis, we now analyze the performance of the proposed scheme and discuss the key parameters affecting the performance. Some global simulation parameters are assumed as follows: the wavelength of the seed laser $\lambda_0 = 1550 $ nm, the fiber attenuation coefficient $\alpha = -0.2$ dB/km, $\Omega = 15$ GHz, $\eta_D = -0.2$ dB, and the transmission loss of the output coupling mirror for the seed light $T_k = k\tau = 0.05$\cite{Yang13JOSAB, Liu20APS}. Here, we only consider using the high- frequency sidebands of the generated optical frequency comb, as shown in Fig.~\ref{Fig2}. Therefore, the influence of the additional optical noise of the seed laser and the jitter of the OPO cavity length on the system is ignored.

\begin{figure}[t]
\centering
\subfigure[]{
\includegraphics[width=0.47\linewidth]{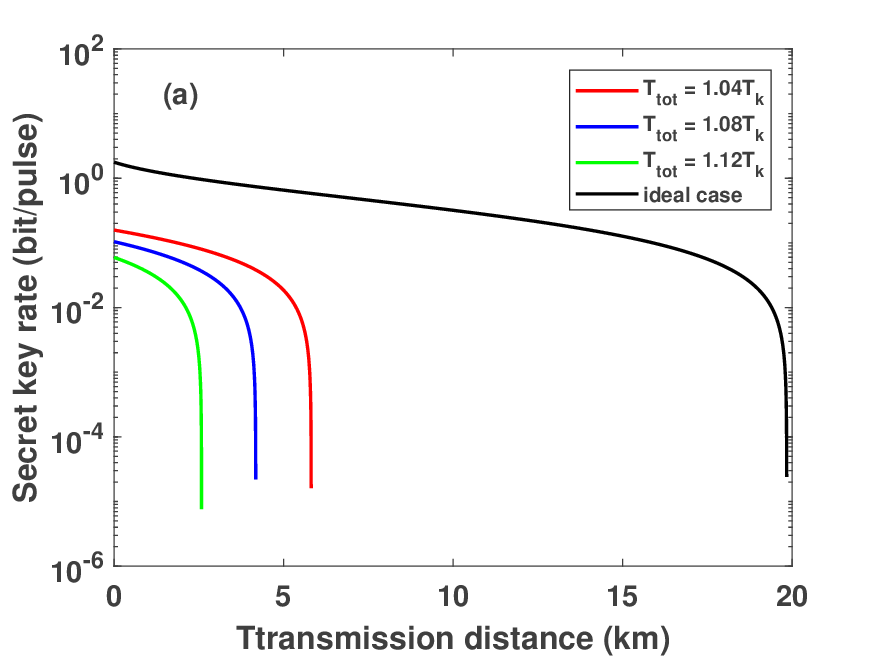}
\label{Key_T_tot1}
}
\subfigure[]{
\centering
\includegraphics[width=0.47\linewidth]{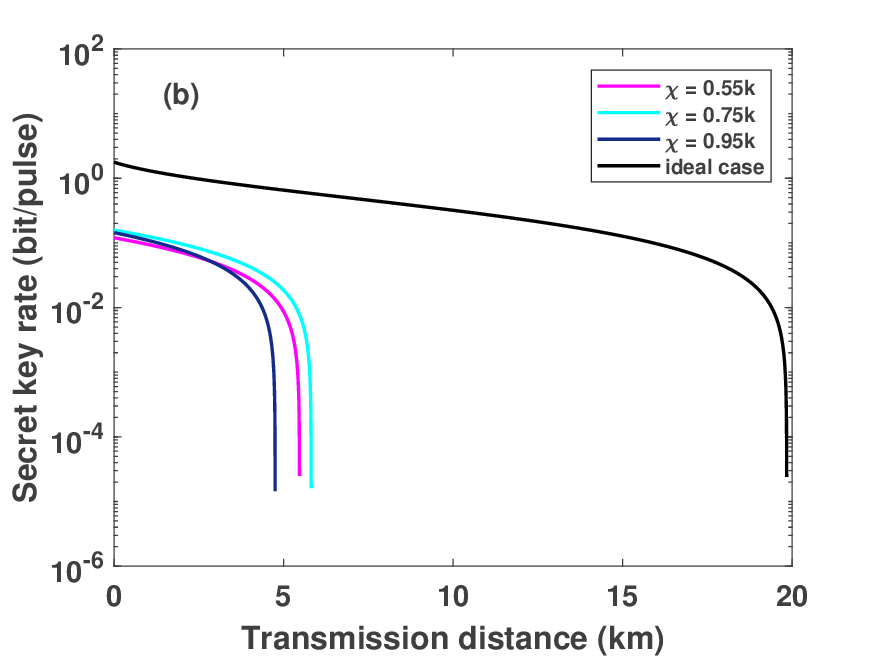}
\label{Key_chi1}
}
\caption{The asymptotical security of the proposed scheme between any two arbitrary users under different OPO parameters. (a) The secret key rate as a function of the transmittance distance for the proposed scheme under different total decay rate $T_{tot}$ of the NOPA. (b) The secret key rate as a function of the transmittance distance for the proposed scheme under different single-pass parametric gain amplitude $\chi$. (c) The secret key rate as a function of the transmittance distance for the proposed scheme under different attenuations $\eta_1$ and $\eta_2$ of the center node. (d) The secret key rate as a function of the transmittance distance for the proposed scheme under different detector parameters.}
\label{Key1}
\end{figure}

First, let's look at the influence of the relevant parameters of the OPO process on the performance of the scheme. We found that the factors affecting the performance of the scheme mainly include two aspects, namely the extra intracavity loss rate of the NOPA and the parametric amplitude gain. Fig.~\ref{Key1} shows the performance of the scheme under different extra intracavity loss rates and parametric amplitude gains of the NOPA. The black solid line represents the ideal situation, that is, $\beta = 0.98$, $T_{tot} = T_k$, $\varepsilon_1 = \varepsilon_2 = 0.01$, $\eta = \eta_1 = \eta_2 = \eta_D = 1$, $\upsilon_{el} = 0$. Fig~\ref{Key1} (a) shows the relationship between the performance of the proposed scheme and $T_{tot}$, where we assume that $\beta = 0.98$, $\chi = 0.75k$, $\eta_1 = \eta_2 = -0.5$ dB, $\eta = 0.9$, $\upsilon_{el} = 0.01$. It can be seen from the figure that the performance of the scheme is very sensitive to the change of $T_{tot}$. When $T_{tot}$ only increases by $4\%T_k$, the performance decreases significantly. Fig~\ref{Key1} (b) shows the change of the performance of the proposed scheme with respect to $\chi$, where most of the parameter settings are the same as those in Fig.~\ref{Key1} (a), but $T_{tot} = 1.04T_k$. The results show that the ratio of $\chi$ and $k$ also has a certain influence on the performance, and there is an optimal $\chi$ that makes the system performance optimal.

\begin{figure}[t]
\centering
\subfigure[]{
\includegraphics[width=0.47\linewidth]{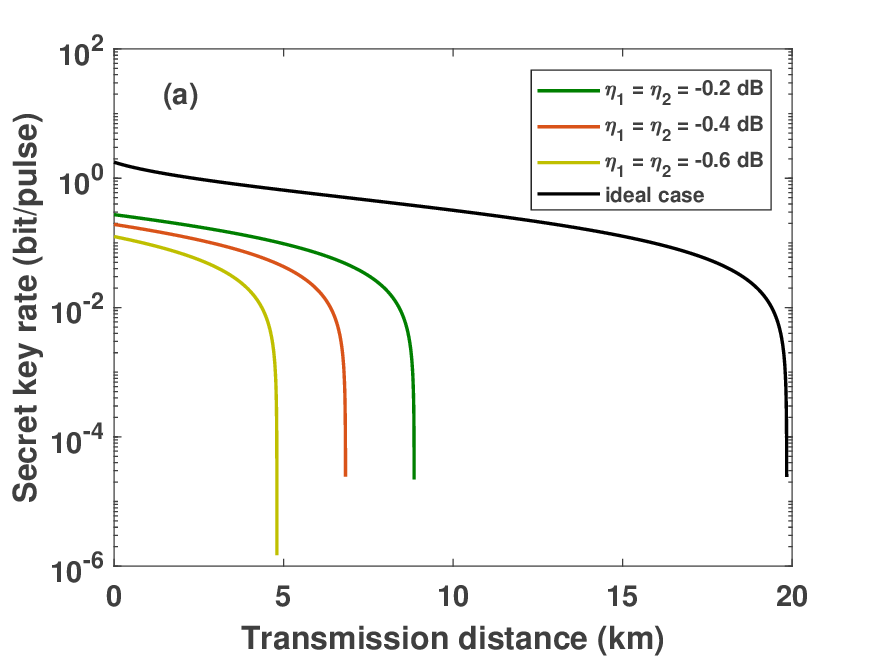}
\label{Key_T_source1}
}
\subfigure[]{
\centering
\includegraphics[width=0.47\linewidth]{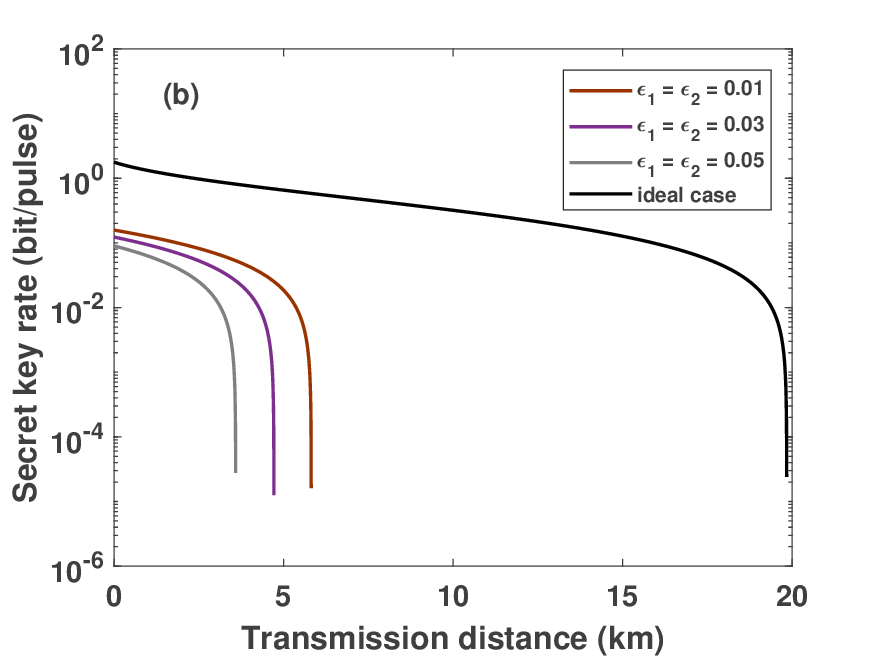}
\label{Key_excess_noise1}
}
\subfigure[]{
\centering
\includegraphics[width=0.47\linewidth]{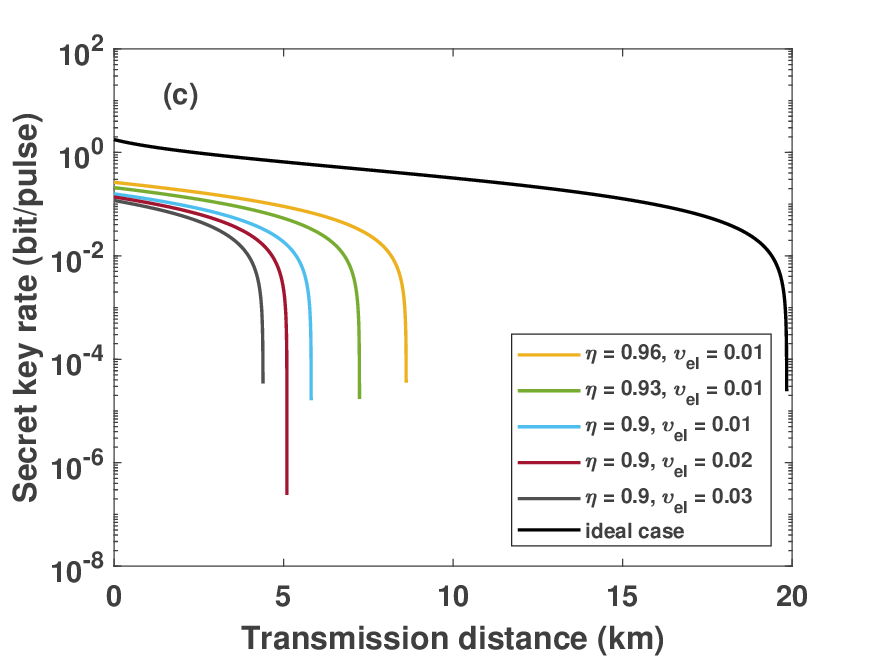}
\label{Key_eta_vel1}
}
\subfigure[]{
\centering
\includegraphics[width=0.47\linewidth]{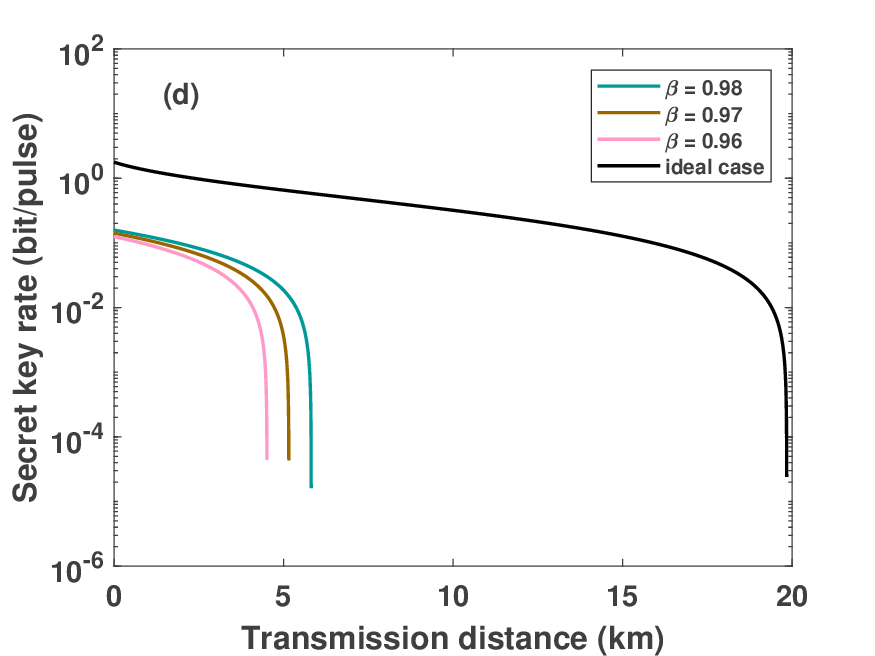}
\label{Key_beta1}
}
\caption{The asymptotical security of the proposed scheme between arbitrary two users under different settings of the system's parameters. (a) The secret key rate as a function of the transmittance distance for the proposed scheme under different values of insertion loss $\eta_1$ and $\eta_2$ of the optical elements in the center node. (b) The secret key rate as a function of the transmittance distance for the proposed scheme under different values of excess noise. (c) The secret key rate as a function of the transmittance distance for the proposed scheme under different detector parameters. (d) The secret key rate as a function of the transmittance distance for the proposed scheme under different reconciliation efficiencies.}
\label{Key2}
\end{figure}

Next, we will analyze the influence of other system parameters on the performance of the scheme, including the transmittances $\eta_1$ and $\eta_2$ of the optical elements at the central node, the excess noises $\varepsilon_1$ and $\varepsilon_2$ of the system, the detection efficiency $\eta$ and electrical noise $\upsilon_{el}$ of the detector, and the reconciliation efficiency $\beta$. Fig.~\ref{Key2} shows the simulation results when the above parameters take different values. The black solid line also represents the ideal situation. The settings of other parameters in each figure are the same except for the changing parameters. Fig.~\ref{Key2} (a) shows the performance of the scheme when $\eta_1$ and $\eta_2$ vary. The settings of parameters other than $\eta_1$ and $\eta_2$ are the same as those in Fig.\ref{Key1}, and we assume that the attenuations of the two modes of TMSS are equal before being sent into the channel. From the results in Fig.~\ref{Key2} (a), it can be concluded that the performance of the system is relatively sensitive to the insertion loss of the optical elements at the central node. A small change in the insertion loss can significantly change the performance. Fig.~\ref{Key2} (b) shows the impact of excess noise on system performance. Obviously, the larger the excess noise, the worse the system performance. Fig.~\ref{Key2} (c) represents the secret key rate versus distance for different detection efficiencies and electrical noise of the detector. The results show that the defects of the detector also have a significant impact on the system performance, especially the efficiency of the detector. Fig.~\ref{Key2} (d) shows the impact of reconciliation efficiency on system performance. The results show that the lower the reconciliation efficiency, the worse the system performance.

According to the above simulation results, we find that in the CVQKD scheme with the entangled source in the middle, the multi-user fully connected CVQKD network based on the optical frequency comb entangled state is feasible for short-distance deployment under well loss and noise control. Multiple factors affect the performance of the proposed CVQKD network system. First, among the OPO process parameters, the extra intracavity loss of the NOPA has a huge impact on the system. A small change in the extra loss will bring about a large performance change. Therefore, the control of the intracavity extra loss is very important in practical applications. Second, the insertion losses of various optical components also have a significant impact on the system. These optical components include fiber couplers, AMs, PBSs, waveshapers, polarization controllers, homodyne detectors, etc. Therefore, it is necessary to minimize the number of optical components and reduce their insertion losses. Third, it is necessary to control the system noise as much as possible, including various technical noises and detector electrical noises. Finally, data post-processing parameters, such as reconciliation efficiency, also have a certain impact on the system performance. In general, various losses in the system are the key factors affecting the network performance.

In terms of the current actual situation, the implementation of a multi-user fully connected CVQKD network based on optical frequency comb entangled states still needs to overcome many problems, especially the losses in various aspects of the system. First of all, the extra intracavity loss of the NOPA still needs to be further reduced experimentally. The extra intracavity loss achieved in the literature~\cite{Liu20APS} is about 1.9\%, which is far from meeting the requirements for network deployment. How to control the extra intracavity loss within 0.5\% or even lower requires breakthroughs in manufacturing technology. Secondly, the insertion losses of the optical components used currently hardly meet the requirements of the proposed scheme. For example, the insertion losses of the AM and the waveshaper in the central node are generally around 2 - 3 dB (such as the WaveShaper 16000). Although their insertion losses have been reduced to a relatively low level, they still need to be further reduced or replaced by other technique. Also, it still requires further efforts to increase the efficiency of the detector (including the losses in the receiving optical path) to 0.9 or higher (usually around 0.6 - 0.7).
Finally, the number of available signal-idler entangled pairs in the system is not only limited by the frequency comb but also affected by the bandwidth of the generated LO and the minimum frequency interval of the wavelength demultiplexing of the waveshaper. The modulation bandwidth of carrier-suppressed double-sideband modulation is generally several tens of GHz. If the free spectral range of the NOPA is 15 GHz and the bandwidth of double-sideband modulation is 45 GHz, then only 6 LO sidebands can be generated, and only 6 entangled pairs can be used in the CVQKD network. The minimum frequency interval of the wavelength demultiplexing of the waveshaper limits the minimum value of the free spectral range. That is to say, the size of the free spectral range must be greater than or equal to the minimum frequency interval. Therefore, it is necessary to select an MZM with the largest possible modulation bandwidth and a waveshaper with the smallest possible minimum frequency interval for wavelength division multiplexing and demultiplexing.

\section{Conclusion}\label{conclusion}
Facing the future large-scale deployment of quantum communication networks, this paper focuses on the CVQKD networking technology and proposes a CVQKD network scheme based on the optical frequency comb entangled state. The proposed scheme utilizes the optical frequency comb formed by the signal-idler sideband entangled pairs generated by the type-$\mathrm{II}$ NOPA, and combines the CVQKD protocol with the entangled source in the middle to implement a CVQKD network with a star topology between the central node and users. Thus a fully connected topology among all users is possible, enabling multiple users to perform CVQKD simultaneously and independently. Simulation results show that a short-distance multi-user fully connected CVQKD network based on the optical frequency comb entangled state using high-frequency optical frequency comb sidebands is feasible. Losses in various aspects of the system are the key factors affecting the system's performance. The suggestions in this paper provide an effective solution for the future development of multi-user CVQKD networks.

\section*{Acknowledgment}
This work was supported by the Scientific Research Fund of Hunan Provincial Education Department (Grant Nos. 24C0168, 24C0165) and the National Natural Science Foundation of China (Grant No. 62501084).

\section*{Disclosures}
The authors declare no conflicts of interest.

\section*{Data availability}
Data underlying the results presented in this paper are not publicly available at this time but may be obtained from the authors upon reasonable request.

\bibliography{sample}






\end{document}